\documentclass{article}
\usepackage{ijcai24}

\usepackage{times}
\usepackage{soul}
\usepackage{url}
\usepackage[hidelinks]{hyperref}
\usepackage[utf8]{inputenc}
\usepackage[small]{caption}
\usepackage{graphicx}
\usepackage{amsmath}
\usepackage{amsthm}
\usepackage{amssymb}
\usepackage{booktabs}
\usepackage{algorithm}
\usepackage{algorithmic}
\usepackage[switch]{lineno}
\usepackage{multirow}
\usepackage{diagbox}
\usepackage{wrapfig}

\title{Leveraging Contrastive Learning for Few-shot Geolocation of Social Posts}

\author{
Menglin Li
\and
Kwan Hui Lim
\affiliations
Singapore University of Technology and Design\\
\emails
menglin\_li@mymail.sutd.edu.sg, kwanhui\_lim@sutd.edu.sg
}

\begin{document}

\maketitle

\begin{abstract}
Social geolocation is an important problem of predicting the originating locations of social media posts. However, this task is challenging due to the need for a substantial volume of training data, alongside well-annotated labels. These issues are further exacerbated by new or less popular locations with insufficient labels, further leading to an imbalanced dataset.
In this paper, we propose \textbf{ContrastGeo}, a \textbf{Contrast}ive learning enhanced framework for few-shot social \textbf{Geo}location.
Specifically, a Tweet-Location Contrastive learning objective is introduced to align representations of tweets and locations within tweet-location pairs.
To capture the correlations between tweets and locations, a Tweet-Location Matching objective is further adopted into the framework and refined via an online hard negative mining approach.
We also develop three fusion strategies with various fusion encoders to better generate joint representations of tweets and locations.
Comprehensive experiments on three social media datasets highlight ContrastGeo's superior performance over several state-of-the-art baselines in few-shot social geolocation.
\end{abstract}

\section{Introduction}
Social geolocation is an important task of estimating the originating locations of individual social media posts, with numerous applications for a host of location-based services, including local recommendations \cite{ho2022poibert,liu2020strategic}, location-based advertisements \cite{huang2018location,evans2012intelligent}, emergency location identification \cite{scalia2022cime}, and disaster management \cite{zheng2018survey}.

\begin{wrapfigure}{r}{4.3cm}
  \centering
\includegraphics[width=4.3cm, trim=12mm 0mm 0mm 0mm]{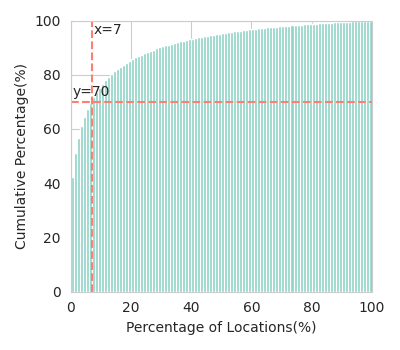}
  \caption{Cumulative distribution of tweet counts against locations from the Twitter-SG dataset.}
  \label{fig:distribution}
\end{wrapfigure}

Social geolocation is typically formulated as a classification problem, frequently solved using supervised models \cite{li2018location}. 
While this approach is effective, there are numerous challenges, such as: (i) the need for a substantial volume of training data, alongside well-annotated labels for optimal performance; (ii) the evolving characteristics of locations (e.g., new businesses) due to factors like economic development, transportation infrastructure changes, climate fluctuations, and even events like the COVID-19 pandemic \cite{li2021disparate}; (iii) the data imbalance problem, as illustrated in Figure \ref{fig:distribution}, where a small percentage (7\%) of locations contain the majority (70\%) of visits/tweets, and many unpopular locations have insufficient labels.

Contrastive learning, which aims to minimize (maximize) the distance between positive (negative) item pairs, has emerged as a powerful paradigm for representation learning and garnered the attention of both academia and industry \cite{khosla2020supervised}. It has recently gained traction across numerous domains such as Natural Language Processing (NLP) with models like SimCSE \cite{gao-etal-2021-simcse} and ConSERT\cite{yan-etal-2021-consert}, as well as Computer Vision (CV) and Multi-modality (MM) with models like MOCO \cite{he2020momentum}, CLIP \cite{radford2021learning}, and SimVLM \cite{wang2022simvlm}, which highlight its effectiveness, especially in the context of open-vocabulary inference, where the data scarcity challenge is more prominent.
This naturally raises the question: Can contrastive learning contribute to the enhancement of social geolocation performance for previously almost "unseen" locations?

Inspired by the intuitions behind contrastive learning, we explore its potential to enhance social geolocation, particularly for locations with sparse data. Our contributions are as follows:
\begin{itemize}
    \item We introduce ContrastGeo, a framework specially designed for few-shot social geolocation. In the initial step, we aggregate pertinent information from tweets and locations separately. Subsequently, this information is fed into a pre-trained language model as an encoder to generate embeddings for subsequent training (Section~\ref{ssec:archi}).\footnote{Codes will be made publicly available after paper acceptance.}
    \item We propose the Tweet-Location Contrastive learning objective (TLC) to align representations of tweets and locations within tweet-location pairs, via the computation of contrastive similarity (Section~\ref{ssec:tlc}).
    \item We also develop an additional loss, the Tweet-Location Matching objective (TLM), designed to capture the correlations of tweets and locations, refined using an online hard negative mining approach based on contrastive similarity (Section~\ref{ssec:tlm}).
    \item We develop three designs of the fusion module incorporating various fusion encoders to effectively generate joint representations of tweets and locations (Section~\ref{ssec:fusion}).
    \item We conduct extensive experiments on three datasets to demonstrate the superior performance of ContrastGeo compared to several state-of-the-art baselines in few-shot social geolocation (Section~\ref{ssec:expResults}), as well as a comprehensive ablation study to investigate the effects of model architecture, hard negatives, pooling methods, fusion types, prompt design, and temperature on ContrastGeo (Section~\ref{ssec:ablationStudy}).
\end{itemize}

\begin{figure*}[!htb]
  \centering
  \includegraphics[width=0.9\linewidth]{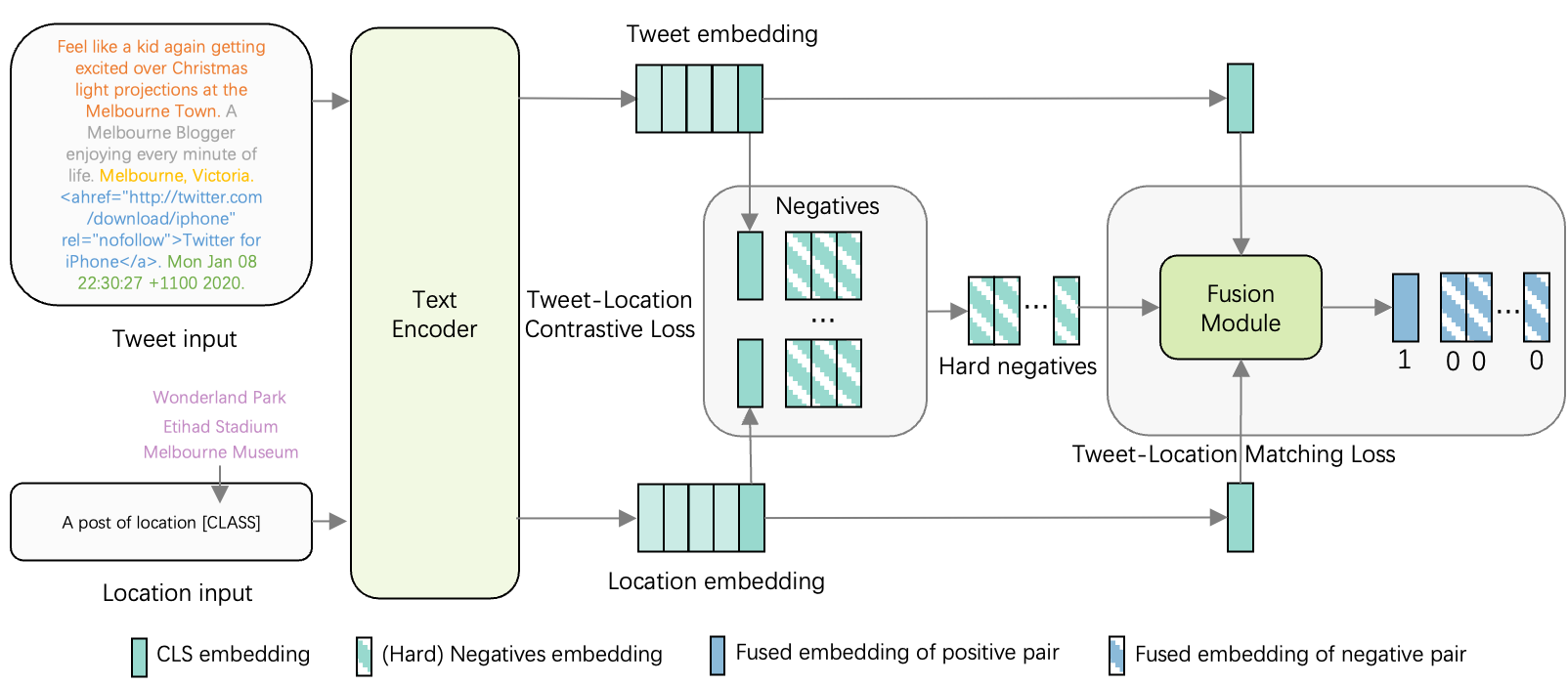}
  \caption{Model Architecture of ContrastGeo.}
  \label{fig:diagram}
\end{figure*}

\section{Related Work}
\subsection{Contrastive Learning}


Contrastive learning has emerged as a powerful paradigm for representation learning, achieving state-of-the-art performance across multiple domains like NLP, CV, and MM. 
Its underlying principle is based on aligning positive pairs while diverging negative pairs, and many studies utilized it for pre-training in a self-supervised manner with large volumes of unlabeled data, like in MoCo~\cite{he2020momentum}, SimCLR~\cite{chen2020simple} and SimCSE~\cite{gao-etal-2021-simcse}.
In contrast, other models like CLIP~\cite{radford2021learning} and ALBEF~\cite{li2021align} utilize a supervised learning approach to contrastive learning. For example, CLIP uses descriptive text as a supervisory signal during pre-training to develop versatile visual models that are highly effective in zero-shot and few-shot settings. Due to its flexibility, high applicability, and adversarial robustness, this approach have led to numerous visual and linguistic models, such as CLIPasso~\cite{vinker2022clipasso}, ActionCLIP~\cite{wang2021actionclip}, CLIP4CLIP~\cite{luo2022clip4clip} and PointCLIP~\cite{zhang2022pointclip}.

\subsection{Social Geolocation}
Social geolocation aims to infer locations from social media data, and we focus on estimating the originating locations of individual posts in this work.
Researchers have employ diverse methodologies ranging from probabilistic language models \cite{ozdikis2018spatial} and convolutional networks \cite{iso2017density} to advanced techniques like BERT models \cite{scherrer2021social} and neural networks \cite{wang2021semantic}, often integrating metadata for enhanced accuracy \cite{mircea2020real}.
These approaches are tailored for various scenarios, including disaster events \cite{singh2019event,ouaret2019random}, and leverage features like term co-occurrences, user history, and multi-level geocoding \cite{kulkarni2020spatial} to accurately infer locations from social posts.

\subsubsection{Social Geolocation with Contrastive Learning}\label{sssec:geo-con}
The success of contrastive learning in other domains has prompted researchers to explore its application in social geolocation. 
A study focused on geolocating Italian social media posts employed contrastive learning with data augmentation, creating positive and negative tweet pairs based on regional dialect differences \cite{koudounas2023barhotti}.
A two-stage framework, consisting of a city classifier and a place identifier, is proposed for detecting points of interest through the combined use of images and text from social networks    \cite{lucas2022detecting}.
The place classifier utilizes the transfer learning capabilities of CLIP to perform zero-shot multimodal geolocation.
Contrastive learning is incorporated into some related tasks, including street-level IP geolocation \cite{tai2023ripgeo}, next POI recommendation \cite{oh2023pre}, and geographical representation learning \cite{fang2023representing,huang2021m3g,bai2023geographic}.

\subsubsection{Few-Shot Social Geolocation}
Addressing the challenge of limited training samples in social geolocation, some researchers have explored meta-learning approaches, which quickly adapt to new user locations \cite{zhou2022metageo}, and BERT models with specialized training dataset constructions \cite{suwaileh2022disaster} for zero-shot or few-shot settings. 
Episodic learning has also been proposed to tackle population imbalance issues in user geolocation, extracting a single representation for each class to enable balanced class distribution training \cite{tao2021episodic}.

\section{Proposed Method}\label{sec:method}

\subsection{Background: Contrastive Learning}\label{ssec:bg}
Inspired by the impressive zero-shot and few-shot performance of CLIP and its variants, we initially considered leveraging CLIP for geolocation. 
However, CLIP thrives in multi-modal environments but social platforms predominantly feature text, which severely limits the applicability of CLIP in this context. Furthermore, this mismatch leads to issues such as the need for image data generation or a decrease in generalizability, as observed by \cite{lucas2022detecting}.



We then turn to the idea of contrasting samples (tweets/images) with categories (locations/texts) in a supervised setting to leverage upon the learning capabilities of CLIP. This approach poses a significant challenge due to the inherent discrepancy in the number of categories and samples, i.e., more than one sample may belong to the same class in an input batch. Most conventional contrastive loss, including that of CLIP, are unable to handle this situation of constructing in-batch labels. 
Researchers have proposed contrasting samples by forming positive and negative pairs based on categories~\cite{khosla2020supervised}, and utilizing external dialect corpora to create paired samples~\cite{lucas2022detecting}. However, these approaches require complex data augmentation, a lengthy two-stage training process, and additional sometimes inaccessible resources.

In contrast, our research proposes to directly contrast tweets with their respective locations, employing a dual-objective system, contrastive loss coupled with matching loss, in an end-to-end training framework. 
This method efficiently adapts to varying numbers of positive and hard negative samples, leveraging available labels.

\subsection{Model Architecture}\label{ssec:archi}
Our proposed ContrastGeo comprises a text encoder, a fusion module, and two training objectives: Tweet-Location Contrastive learning (TLC) and Tweet-Location Matching loss (TLM). Fig. \ref{fig:diagram} shows the architecture of ContrastGeo.

ContrastGeo takes both tweet and location data as input.
For tweet input, relevant tweet attributes (i.e., tweet contents, users' descriptions, users' hometowns, posting sources, and timestamps) are concatenated to an input sentences. 
For location input, we insert $K$ category names (locations) into the class token slot of a predefined template: "a post of location [CLASS]".
Subsequently, both tweet and location inputs are encoded through a pre-trained language model, like BERT. 

Prior to fusion, we apply TLC to fine-tune the text encoder, which aligns the representations of tweets and locations within tweet-location pairs.
Following this, We employ TLM, which leverages global hard negatives identified via contrastive similarity, to facilitate the learning of tweet-location correlations using a fusion module.
ContrastGeo is trained by employing the dual-objective strategy.

\subsection{Tweet-Location Contrastive Learning}\label{ssec:tlc}
In the domain of social geolocation, efficient representations of tweets and locations are critical for capturing the underlying correlations between these entities. 
To this end, we propose Tweet-Location Contrastive learning (TLC) to learn representations from positive tweet-location pairs.
Within a given batch, it is not uncommon to observe multiple tweets that are annotated with identical location tags.
To handle such an issue of in-batch labels, our approach contrasts each tweet in the input batch directly with all locations from the dataset.
A tweet-location pair is deemed positive if the location corresponds to the ground-truth label of the tweet; conversely, it is considered negative if there is no such correspondence.
Since the total count of unique locations is generally finite and manageable, this strategy does not incur much computational overhead. 
It is notable that for each input batch of tweets, we contrast it with the same full label set.

Consider a batch consisting of \(N\) tweet sentences, \(\{\mathbf{t}_i\}_{i=1, ..., N}\), randomly sampled from the dataset.
Correspondingly, we identify a label set of all \(K\) unique locations from the dataset, represented as \(\{\mathbf{l}_j\}_{j=1, ..., K}\).
The text encoder maps an input tweet \(\mathbf{t}\) to a sequence of embedding vectors \(\{\mathbf{e}_{cls}, \mathbf{e}_1, ..., \mathbf{e}_m\}\).

Here, \(f_{\theta}(\mathbf{t})=\mathbf{e}_{cls}\) is utilized to signify the sentence-level embedding of a tweet \(\mathbf{t}\).
Analogously, \(f_{\theta}(\mathbf{l})\) represents the encoded sentence embedding for a prompted location sentence~\(\mathbf{l}\).
For each pair of tweet and location, we compute a softmax-normalized similarity as follows:
\begin{equation}\label{eq:sim}
    p_{tlc}(\mathbf{t}_i, \mathbf{l}_j) = \frac{exp(sim(f_{\theta}(\mathbf{t}_i), f_{\theta}(\mathbf{l}_j))/\tau)}{\sum_{j=1}^{K}exp(sim(f_{\theta}(\mathbf{t}_i), f_{\theta}(\mathbf{l}_j))/\tau)},
\end{equation}
where \(\tau\) denotes the temperature scaling parameter and \(sim(\mathbf{t}, \mathbf{l})\) is the cosine similarity function defined as \(\frac{\mathbf{t}^{T} \mathbf{l} }{\|\mathbf{t}\| \cdot \|\mathbf{l}\|}\).
Let \(\mathbf{y}_{tlc}(t, l)\) be the one-hot encoded ground-truth similarity vector, with a value of 1 assigned to positive tweet-location pairs and 0 otherwise.
The contrastive loss function for tweet-location pairs is thus formalized as the cross-entropy \(F_{cross-entropy}\) between the predicted probabilities \(\mathbf{p}_{tlc}\) and the true distributions \(\mathbf{y}_{tlc}\):
\begin{equation}\label{eq:tlc}
    \mathcal{L}_{tlc} = \mathbb{E}_{(\mathbf{t}, \mathbf{l}) \sim \mathcal{D}} [F_{cross-entropy}(\mathbf{y}_{tlc}(t, l), \mathbf{p}_{tlc}(t, l))].
\end{equation}

This formulation aims to optimize the representation of tweets and locations by minimizing the contrastive loss, thereby improving the efficacy of social geolocation tasks.

\subsection{Tweet-Location Matching}\label{ssec:tlm}
Taking inspiration from the ALBEF framework \cite{li2021align}, our training regimen incorporates an additional objective: a Tweet-Location Matching loss (TLM), which is bolstered by an approach of hard negative mining, to learn and exploit the intricate dynamics between tweets and locations in a joint training paradigm. 
Furthermore, TLM is modified to accommodate an arbitrary number of negative pairs.

In our setup, a tweet and its ground-truth location (matched) constitute a positive pair. 
Compared with the TLC objective, the TLM objective requires an explicit construction of negative pairs. 
To elevate the effectiveness of the learning process, we adopt a strategy that identifies and leverages multiple hard negatives—those tweet-location negative pairs that, despite their semantic proximity, differ in subtler, more specific aspects. 
We find these hard negatives without incurring additional computational costs, utilizing the contrastive similarity delineated in Equation \ref{eq:sim} as our selection criterion.

For each tweet in a given batch, we embark on a mining operation to extract $M$ most informative negative locations from the total set of $K$ unique locations excluding the ground-truth location.
We present two methodologies to extract these hard negatives. One approach samples $M$ negative locations from a multinominal distribution weighted by contrastive similarity, thus preferentially selecting those locations with higher resemblance to the tweet. The other approach directly identifies the top-$M$ locations with the highest contrastive similarity scores.
We denote these two approaches as \textit{multinominal} and \textit{top}, respectively.

Thereafter, the obtained negative locations are coupled with the tweet to form $M$ negative pairs.
These, alongside the positive pair, are processed by the fusion encoder to generate the joint representation for each tweet-location pair, the design of which we elaborate in Section \ref{ssec:fusion}.
The representations of one positive and $M$ negative pairs are subsequently arranged into a $(M+1)$-dimensional vector and presented to a classification head. 
This classifier, comprising a fully connected (FC) layer followed by a softmax activation, outputs the probability distribution \(\mathbf{p}_{tlm}\) over the $(M+1)$ classes.

The TLM loss is thus articulated as the expected cross-entropy loss between the ground truth distribution \(\mathbf{y}_{tlm}(t, l)\) and the predicted distribution \(\mathbf{p}_{tlm}\), as shown in Equation \ref{eq:tlm}.
\(\mathbf{y}_{tlm}(t, l)\) is a  \((M+1)\)-dimensional vector, wherein the value of the position pair (as the first element of the vector) is assigned as 1 and all negative pairs 0.
\begin{equation}\label{eq:tlm}
    \mathcal{L}_{tlm} = \mathbb{E}_{(\mathbf{t}, \mathbf{l}) \sim \mathcal{D}} [F_{cross-entropy}(\mathbf{y}_{tlm}(t, l), \mathbf{p}_{tlm}(t, l))]
\end{equation}
The overall training objective for ContrastGeo integrates the TLC and TLM losses, that is:
\begin{equation}
    \mathcal{L} = \mathcal{L}_{tlc} + \mathcal{L}_{tlm}. 
\end{equation}
This comprehensive objective propels ContrastGeo to better discern and internalize the nuanced associations between tweets and their respective geographic knowledge.

\subsection{Fusion Module}\label{ssec:fusion}

\begin{figure}[!htb]
  \centering
  \includegraphics[width=\linewidth]{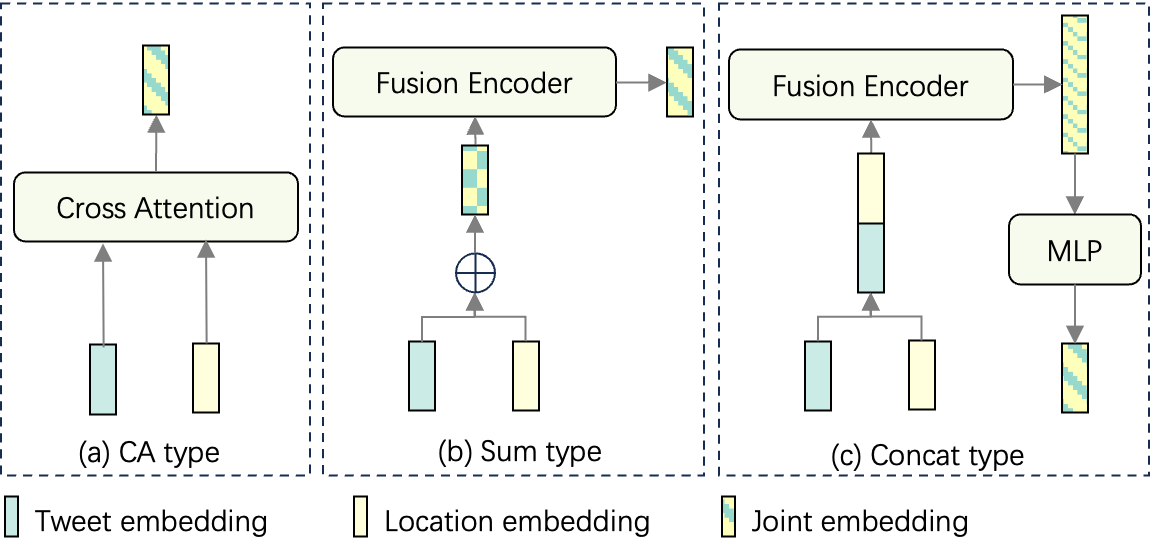}
  \caption{Three types of the fusion module. For Sum type and Concat type, the fusion encoder can be Multi-Head Attention (MHA), a Bottle-Neck Adapter (BNA), Transformer Encoder (TE), and Multi-Layer Perceptron (MLP).
  }
  \label{fig:fusion}
\end{figure}

In the process of calculating TLM, a critical step involves the fusion of tweet and location embeddings. 
Our approach is grounded in a pre-trained language model, necessitating the judicious integration of new learnable weights within the fusion module. 
The absence of proper weight initialization can impede learning and adversely affect the performance of the pre-trained model during backpropagation training. 
Consequently, we classify the fusion module mechanisms into three distinct categories, each characterized by its method of combining tweet and location embeddings. 
Figure \ref{fig:fusion} delineates the architecture of these three mechanisms.

\begin{table}[!htb]
  \caption{Statistics of Experimental Datasets.}
  \label{tab:dataset}
  \centering
  \resizebox{0.43\textwidth}{!}{
  \begin{tabular}{cccc}
    \toprule
    Dataset & Categories & Dev Samples & Test Samples\\
    \midrule
    Twitter-Mel     & 108 & 5,033 & 32,669\\
    Flickr-Mel     & 45 & 1,465 & 9,517\\
    Twitter-SG     & 199 & 4,968 & 32,256\\
  \bottomrule
\end{tabular}}
\end{table}


\vspace{2.3mm}
{\noindent\bf CA Type.}
Drawing inspiration from ALBEF \cite{li2021align}, which incorporates cross-attention into transformer layers for feature fusion, our CA type mechanism employs cross-attention to derive joint representations of tweets and locations while minimizing parameter involvement. 
Specifically, in the cross-attention computation, tweet embeddings serve as Query, and location embeddings function as Key and Value, directly yielding the joint representations.

\begin{figure*}[!htb]
  \centering
  \includegraphics[width=\linewidth]{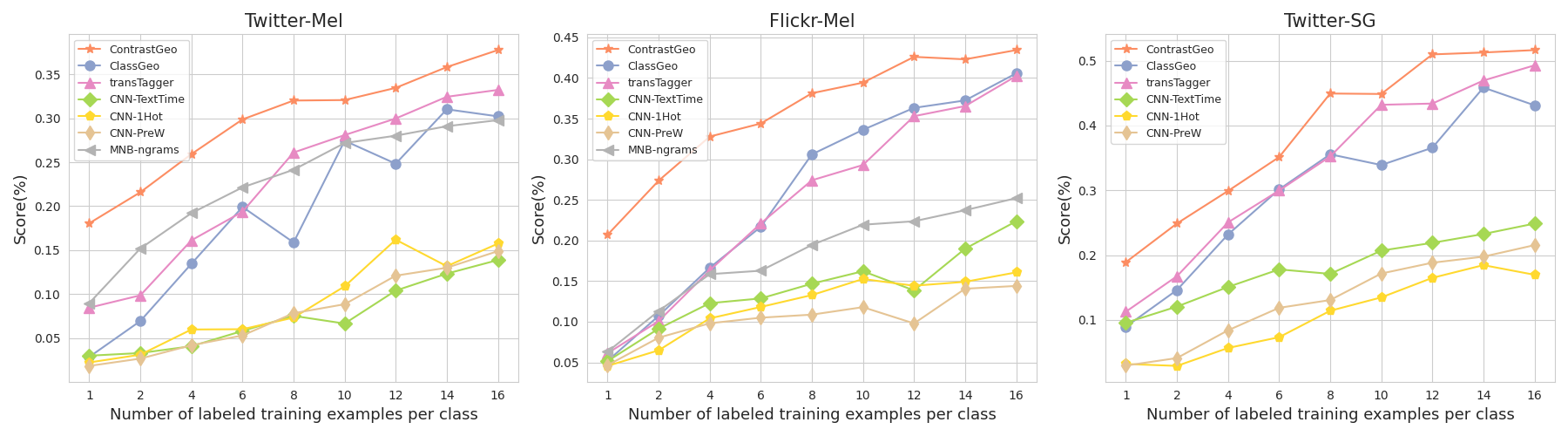}
  \caption{Few-shot performance comparison between ContrastGeo, ClassGeo, and representative geolocation models, on Twitter-Mel, Flickr-Mel, Twitter-SG in terms of $accuracy$. Our ContrastGeo shows consistent superiority to other models under 1, 2, 4, 6, 8, 10, 12, 14, 16-shot settings.}
  \label{fig:fsl}
\end{figure*}

\vspace{2.3mm}
{\noindent\bf Sum Type.}
Element-wise addition and concatenation are popular and fundamental techniques for feature fusion, underpinning the Sum and Concat types, respectively. 
In the Sum type, we first perform an element-wise addition of tweet and location embeddings. 
As this alone is insufficient for a robust joint representation, we introduce a fusion encoder to enhance the capture of tweet-location correlations. 
Various structural options exist for this encoder, including the inter-view adapter in PointCLIP \cite{zhang2022pointclip} that is akin to the bottle-neck adapter \cite{houlsby2019parameter}, a field-level multi-head attention layer used in HLPNN \cite{huang-carley-2019-hierarchical}, and a multi-layer transformer encoder similar to transTagger's approach \cite{li2023transformer}.

We implement our fusion encoder with a Multi-Head Attention (MHA) layer, a Bottle-Neck Adapter (BNA), and a Transformer Encoder (TE) layer. 
Additionally, a Multi-Layer Perceptron (MLP) is employed as an alternate option. 
Each fusion encoder variant produces joint representations that keep dimensions unchanged, facilitating subsequent TLM computations.

\vspace{2.3mm}
{\noindent\bf Concat Type.}
As implied by its name, the Concat type involves the concatenation of tweet and location embeddings. 
This resulting intermediate feature is then processed through a fusion encoder, mirroring the structure outlined for the Sum type. 
Following which, an additional MLP layer is applied to retain the dimensional consistency. 
Our subsequent empirical evaluations (Section \ref{ssec:ablationStudy}) reveal that the simplest configuration—Sum type with an MLP fusion encoder—yields the most effective performance.
This finding aligns with the principle that new learnable weights should be incorporated with caution when working with pre-trained models.

\begin{table*}[!htb]
  \caption{Performance comparison between ContrastGeo against various baseline geolocation models under the 16-shot setting, on Twitter-Mel, Flickr-Mel, Twitter-SG in terms of $accuracy$, $meanDist$, $medDist$. Our ContrastGeo surpasses other models across all metrics.}
  \label{tab:16shot}
  \centering
  \resizebox{1\textwidth}{!}{
  \begin{tabular}{lccccccccc}
    \toprule
    & \multicolumn{3}{c}{Twitter-Mel} & \multicolumn{3}{c}{Flickr-Mel} & \multicolumn{3}{c}{Twitter-SG} \\ \cline{2-10}
 Model & $accuracy\uparrow$ & $meanDist\downarrow$ & $medDist\downarrow$ & $accuracy\uparrow$ & $meanDist\downarrow$ & $medDist\downarrow$  &$accuracy\uparrow$ & $meanDist\downarrow$ & $medDist\downarrow$  \\
    \midrule
    transTagger	& 33.2 	& 816.6 	& 442.0 	& 40.2 	& 657.4 	& 329.7 	& 49.3 	& 2334.0 	& 57.9 \\
    CNN-TextTime	& 13.9 &  	1329.9 	& 997.6 	& 22.4 	& 1116.9 	& 906.3 	& 24.9 	& 3880.3 	& 1340.9  \\
    CNN-1Hot	& 15.7 	& 1347.9 	& 1145.3 	& 16.1 	& 1199.4 	& 1079.0 	& 17.0 	& 4599.8 	& 3569.4  \\
    CNN-PreW	& 14.9 	& 1291.9 	& 1047.4 	& 14.4 	& 1177.2 	& 984.9 	& 21.6 	& 5418.5 	& 2037.3  \\
    MNB-ngrams	& 29.8 	& 1029.9 	& 647.3 	& 25.2 	& 1033.0 	& 800.0 	& 0.2 	& 6715.8 	& 3771.7  \\
    ClassGeo	& 30.2 	& 843.8 	& 518.3 	& 40.6 	& 629.0 	& 309.6 	& 43.1 	& 2626.8 	& 362.4  \\
    ContrastGeo	& \textbf{37.8} 	& \textbf{766.0} 	& \textbf{325.6} 	& \textbf{43.4} 	& \textbf{571.8} 	& \textbf{161.2 }	& \textbf{51.6} 	& \textbf{2267.0} 	& \textbf{0.0}  \\
  \bottomrule
\end{tabular}}
\end{table*}

\section{Experiments}
\subsection{Evaluation Setting}
\vspace{2.3mm}
{\noindent\bf Datasets.}
We performed our experiments on three public datasets, Twitter-Mel, Flickr-Mel, and Twitter-SG, comprising 827K social media posts and their geotagged locations across two cities (Melbourne and Singapore) from Twitter/X and Flickr~\cite{li2023transformer}. 
For more details, we refer readers to~\cite{li2023transformer}.
To adapt these datasets for few-shot social geolocation, we filter out categories and their corresponding samples due to insufficient quantities. 
For each category, we extract all samples and perform a random split into training, development, and test subsets using a ratio of 8.5:0.2:1.3. 
These subsets are then aggregated to form the corresponding training, development, and test sets. 

For the $S$-shot setting, we randomly select $S$ samples from each category in the training set. 
To ensure a fair assessment of few-shot performance, we generate multiple iterations of the few-shot datasets using different random seeds. 
Each few-shot experiment is executed across three distinct iterations/sub-datasets. 
The average of these evaluations is reported to mitigate the variance introduced by the limited number of training samples.
Table \ref{tab:dataset} presents the statistical breakdown of the datasets.

\vspace{2.3mm}
{\noindent\bf Evaluation Metrics.}
Consistent with the commonly used setting in related research \cite{huang-carley-2019-hierarchical,tao2021episodic}, we evaluate geolocation performance using two types of metrics: accuracy and distance error. 
Accuracy is denoted as $accuracy$. 
Distance error is reported in terms of mean and median values, represented as $meanDist$ and $medDist$, respectively.

\vspace{2.3mm}
{\noindent\bf Training Details.}
Our model, ContrastGeo, is developed using the \textit{transformers} package and initialized from pre-trained BERT(cased) checkpoints. 
We take [CLS] representations as the sentence embeddings (refer to Section~\ref{sssec:pool} for a comparative analysis of different pooling methods). 
The model is trained with the AdamW optimizer and an early-stopping mechanism, while the evaluation is based on the development set of Twitter-Mel. 
We then select the best-performing checkpoint for final evaluation on the test sets. 
Using the 16-shot Twitter-Mel setting, we conducted a comprehensive grid search across various parameters, before selecting a batch size of 8, 100 epochs, learning rate of 2e-5, 160 evaluation steps, and AdamW optimizer with beta1 and beta2 of 0.9 and 0.999, respectively.
Adjustments to the evaluation steps are made in accordance with the number of shots and training datasets.

\subsection{Few-Shot Social Geolocation} \label{ssec:expResults}
\vspace{2.3mm}
{\noindent\bf Settings.}
Our evaluation of ContrastGeo covers a range of few-shot settings, specifically 1, 2, 4, 6, 8, 10, 12, 14, and 16 shots, across three datasets: Twitter-Mel, Flickr-Mel, and Twitter-SG. 
We train the model with both TLC and TLM objectives.
To identify hard negatives, we select the top-7 locations with the highest contrastive similarity scores. 
Both TLC and TLM losses are computed based on the CLS representations derived from BERT-encoded embeddings. 
The Sum-type fusion module with an MLP fusion encoder is employed to generate joint representations. 
For the few-shot experiments, we utilize the prompt "a post of location [CLASS]."

\vspace{2.3mm}
{\noindent\bf Performance.}
We compare ContrastGeo against numerous representative geolocation models: transTagger~\cite{li2023transformer}, CNN-TextTime, CNN-1Hot~\cite{Johnson2015EffectiveUO}, CNN-PreW~\cite{kim-2014-convolutional}, and MNB-Ngrams~\cite{ozdikis2018spatial}. 
Additionally, we introduce a variant of ContrastGeo, termed ClassGeo, which replaces the contrastive learning part with a cross-entropy loss for standard classification in social geolocation while maintaining the rest of the model structure.
Figure \ref{fig:fsl} illustrates the few-shot performance of ContrastGeo in terms of $accuracy$.

The results show that ContrastGeo consistently outperforms the comparative models in the few-shot social geolocation task. 
Notably, ContrastGeo demonstrates a superior performance even when limited samples per category are available. 
For instance, ContrastGeo surpasses transTagger by 9.62\% on Twitter-Mel, by 14.59\% on Flickr-Mel, and by 7.64\% on Twitter-SG. 
As the number of training samples increases, the gap becomes smaller, but ContrastGeo continues to maintain a performance lead.

Furthermore, experiment results show that ContrastGeo exhibits significant improvements over ClassGeo under all few-shot scenarios across the three datasets, underscoring the efficacy of incorporating contrastive learning into social geolocation tasks.

Table \ref{tab:16shot} details a comparative analysis of ContrastGeo and the other models under the 16-shot setting, utilizing metrics including $accuracy$, $meanDist$, and $medDist$. 
This comparison offers a comprehensive perspective on performance, where ContrastGeo notably excels across all metrics by a considerable margin.

\subsection{Ablation Study} \label{ssec:ablationStudy}
We now conduct comprehensive ablation studies to study the impacts of various design elements on our model's performance. 
These elements include model architecture, hard negatives, fusion types, prompt design, pooling methods, and temperature settings. 
The reported results represent the average accuracy over three iterations, all conducted on the Twitter-Mel test set under a 16-shot framework.

\vspace{2.3mm}
{\noindent\bf Model Architecture.}
Table \ref{tab:variants} presents a comparative analysis of ContrastGeo variants to evaluate the impact of the model structure and training techniques. 
In contrastive learning, the dual-encoder framework is commonly utilized to process paired samples $(x, x')$.
This typically involves two independent encoders, $f_{\theta_{1}}$ and $f_{\theta_{2}}$, to accommodate the distinct nature of $(x, x')$.
However, as indicated in Table  \ref{tab:variants}, our experimentation with a dual-encoder framework yielded poor results.
We hypothesize that this performance shortfall arises because representations of $(x, x')$ generated by a single encoder are better aligned semantically than those produced by two encoders.
Additionally, the limited number of location samples in our dataset may adversely affect the training of the location-specific text encoder, consequently impacting the performance of the pre-trained language model, and thus the overall efficacy of ContrastGeo.

An attempt to freeze the text encoder within ContrastGeo further reduced performance, possibly due to the significant disparity between the distribution of our geographically rich experimental data and the pre-training corpus of BERT. 
This finding underscores the necessity of fine-tuning for optimal few-shot social geolocation performance.

A comparison between the first and fourth rows in Table \ref{tab:variants} reveals that training solely with the TLC objective leads to a 1.12\% decline in performance, highlighting the importance of the TLM loss.

To address potential overfitting issues stemming from limited training samples, we incorporated a label-smoothing mechanism into ContrastGeo. 
The observed performance decline upon its removal confirms its beneficial role in enhancing the model's generalization capability.
\begin{table}[!htb]
    \caption{Comparative analysis of ContrastGeo variants. 
    ContrastGeo utilizes the one-encoder structure, as shown in the first row, and here the encoder denotes the text encoder. 
    \textit{One-encoder, frozen}: using the one-encoder framework but freezing the text encoder during training. 
    \textit{One-encoder w/o TLM}: training with TLC only.
    \textit{One-encoder w/o label smoothing}: training without applying label smoothing when calculating TLC loss.
    \textit{One-encoder w/ MLP after fusion}: appending an additional MLP layer upon the fusion module.
    }
    \label{tab:variants}
    \centering
    \resizebox{0.35\textwidth}{!}{
    \begin{tabular}{lc}
    \toprule
     Variants & \textit{accuracy} \\
    \midrule
    One-encoder & \textbf{37.8}\\
    Dual-encoder & 2.0 \\
    One-encoder, frozen & 0.7\\
    One-encoder w/o TLM & 36.7\\
    One-encoder w/o label smoothing & 36.2\\
    \bottomrule
\end{tabular}}
\end{table}

\vspace{2.3mm}
{\noindent\bf Hard Negatives.}
As described in Section \ref{ssec:tlm}, our study proposes two methods for mining hard negatives: \textit{multinomial} and \textit{top}. 
To evaluate their effectiveness, we conduct further tests by varying the number of hard negatives (\(M\)) across a range from 1 to 10, as detailed in Table \ref{tab:hard}. 
Our findings do not conclusively favor either the \textit{multinomial} or \textit{top} approach.
The \textit{top} method with \(M = 7\) yields the optimal results.
\begin{table}[!htb]
    \caption{Ablation studies of various hard negative mining policies.}
    \label{tab:hard}
    \centering
    \resizebox{0.5\textwidth}{!}{
    \begin{tabular}{lcccccccccc}
    \toprule
     & 1 & 2 & 3 & 4 & 5 & 6 & 7 & 8 & 9 & 10 \\
    \midrule
    MN & 36.5 & 36.9 & 36.7 & 37.0 & 36.9 & 36.7 & 36.2 & 37.1 & 37.0 & 36.7 \\
    Top & 37.0 & 36.8 & 36.7 & 37.3 & 36.8 & 36.8 & \textbf{37.8} & 36.7 & 36.8 & 36.5 \\
    \bottomrule
\end{tabular}}
\end{table}

\vspace{2.3mm}
{\noindent\bf Pooling Methods.}\label{sssec:pool}
In this section, we conduct an ablation study to assess the impact of various pooling methods on the calculation of the TLC objective in ContrastGeo under a 16-shot setting. 
The prevalent method for constructing sentence embeddings from BERT is the direct use of the [CLS] token. 
Notably, the original BERT architecture employs an additional MLP layer on the [CLS] representation, which corresponds to the \textit{[CLS]} type in Table \ref{tab:pool}. 
Consequently, we include a variant, \textit{[CLS$_{woM}$]}, in our comparison, which utilizes the raw [CLS] token output without this extra MLP layer.

Literature \cite{li-etal-2020-sentence,reimers-gurevych-2019-sentence} also suggests that average embeddings, particularly those derived from both the first and last layers of pre-trained language models, may yield superior results compared to the [CLS] token method alone. 
To explore this, we examine three additional pooling strategies: \textit{First-last avg.}, \textit{Top2 avg.}, and \textit{Avg}. 
Our findings indicate that methods based on the [CLS] token generally surpass those relying on average embeddings, with the \textit{[CLS]} approach being the most effective. 
Therefore, we base the computation of the TLC loss on [CLS] representations.
\begin{table}[!htb]
    \caption{Ablation studies of different pooling methods.}
    \label{tab:pool}
    \centering
    \resizebox{\columnwidth}{!}{
    \begin{tabular}{lccccc}
    \toprule
     Pooler & {CLS} & {CLS}$_{woM}$ & First-last avg. & Top2 avg. & Avg. \\
    \midrule
    \textit{accuracy} & \textbf{37.8} & 37.3 & 37.2 & 36.9 & 37.6 \\
    \bottomrule
\end{tabular}}
\end{table}

\vspace{2.3mm}
{\noindent\bf Fusion Types.}
In Section \ref{ssec:fusion}, we introduce three distinct fusion module types—CA Type, Sum Type, and Concat Type—paired with four possible fusion encoder options. 
This section explores the impact of these nine fusion module configurations on model performance. 
The results of this investigation are detailed in Table \ref{tab:fusion}. 
Additionally, we incorporate two sentence embedding pooling methods, \textit{[CLS]} and \textit{[CLS$_{woM}$]}, which deliver good performance in Section \ref{sssec:pool}, into our comparative analysis.
Our empirical evaluations indicate that the Sum Type fusion module, combined with an MLP fusion encoder and utilizing \textit{[CLS$_{woM}$]} for sentence embeddings, delivers the most optimal performance. 
\begin{table}[!htb]
    \caption{Ablation studies of different fusion module configurations.}
    \label{tab:fusion}
    \centering
    \resizebox{\columnwidth}{!}{
    \begin{tabular}{lccccccccc}
    \toprule
     & CA Type & \multicolumn{4}{c}{Sum Type} & \multicolumn{4}{c}{Concat Type}  \\ \cline{2-10}
     & & MHA    & BNA   & MLP   & TE    & MHA    & BNA   & MLP   & TE\\
    \midrule
    CLS & 37.0 & 37.2 & 37.3 & 37.2 & 37.1 & 37.5 & 36.9 & 36.8 & 37.2  \\
    CLS$_{woM}$ & 36.8 & 37.0  & 36.9 & \textbf{37.8} & 37.3 & 37.6 & 37.4 & 36.0 & 37.1  \\
    Avg.  & - & - & - & - & - & 36.3 & 37.3 & 36.8 & 37.2  \\
    \bottomrule
\end{tabular}}
\end{table}

\vspace{2.3mm}
{\noindent\bf Prompt Design.}
Table \ref{tab:prompt} provides a detailed overview of various prompt designs experimented with in the 16-shot ContrastGeo framework. 
Our findings indicate that all prompts boost the geolocation performance to varying degrees compared to no prompt.
Moreover, the prompt "a post of location [CLASS]." leads to a notable increase in accuracy, specifically by 1.62\%. 

Considering that the ablation studies are conducted on the Twitter-Mel dataset, we also explored incorporating city-specific information into the prompt, exemplified by the addition of "in Melbourne." 
Surprisingly, this modification resulted in diminished performance. 
This outcome may be attributed to the nature of our task, which focuses on fine-grained social geolocation within a specific city—in this case, Melbourne. 
The inclusion of such a broad geographical context, therefore, may not contribute valuable insights for this fine-grained inference and could potentially introduce confusion or misleading information.
\begin{table}[!htb]
    \caption{Performance of ContrastGeo with different prompt designs. [CLASS] denotes the class token.}
    \label{tab:prompt}
    \centering
    \resizebox{0.45\textwidth}{!}{
    \begin{tabular}{lc}
    \toprule
     Prompts & $accuracy$ \\
    \midrule
    "a post of location [CLASS]." & \textbf{37.8}\\
    "a post of [CLASS]." & 37.1\\
    "a post of location [CLASS], in Melbourne." & 36.6\\
    "This post is about location [CLASS]." & 37.5\\
    "This post is about a location [CLASS]." & 36.6\\
    "This post is about the location [CLASS]." & 37.2\\
    "This post is about the place [CLASS]." & 36.8\\
    "This post is about location [CLASS], in Melbourne." & 36.6\\
    No prompt: "[CLASS]" & 36.1\\
    \bottomrule
\end{tabular}}
\end{table}

\vspace{2.3mm}
{\noindent\bf Temperature.}
Our research includes an ablation study focusing on the temperature parameter (\(\tau\)) in the context of contrastive learning, recognizing its critical role as a hyperparameter. 
The study systematically evaluates various values of \(\tau\) to ascertain its impact on model performance. 
As detailed in Table \ref{tab:temp}, our findings reveal that a temperature setting of 0.05 yields the most favorable results.
\begin{table}[!htb]
    \caption{Ablation studies of temperature values.}
    \label{tab:temp}
    \centering
    \resizebox{0.4\textwidth}{!}{
    \begin{tabular}{lcccccc}
    \toprule
     $\tau $ & 0.01 & 0.03 & 0.05 & 0.07 & 0.1 & 0.3 \\
    \midrule
    \textit{accuracy} & 37.5 & 37.1& \textbf{37.8}& 36.9 & 36.6 & 35.2 \\
    \bottomrule
\end{tabular}}
\end{table}

\section{Conclusion and Discussion}
In this work, we introduce ContrastGeo, a model that significantly enhances few-shot performance in social post geolocation. 
ContrastGeo adeptly bridges the gap between social geolocation and traditional contrastive learning through the implementation of an intermediate TLC objective. 
Additionally, the TLM loss, augmented by hard negative mining based on contrastive similarity, effectively captures the intricate relationships between tweets and locations.
Our comprehensive experimental evaluation demonstrates ContrastGeo's superiority over existing state-of-the-art geolocation models, across a spectrum of metrics, datasets, and few-shot scenarios. 
We further conduct meticulous ablation studies to investigate the impact of key components in ContrastGeo, such as model architecture, hard negatives, and fusion types, among others.
Looking ahead, our future work will extend the applications of contrastive learning beyond social geolocation, exploring its potential across a broader range of social analysis tasks.


\bibliographystyle{named}
\bibliography{ijcai24}

\end{document}